\documentclass[conference]{IEEEtran}
\IEEEoverridecommandlockouts
\usepackage{cite}
\usepackage{amsmath,amssymb,amsfonts}
\usepackage{algorithmic}
\usepackage{graphicx}
\usepackage{array}
\usepackage{adjustbox}
\usepackage{textcomp}
\usepackage{xcolor}
\usepackage{multirow}
\usepackage{url}
\usepackage{hyperref}

\def\BibTeX{{\rm B\kern-.05em{\sc i\kern-.025em b}\kern-.08em
    T\kern-.1667em\lower.7ex\hbox{E}\kern-.125emX}}
\begin{document}

\title{FAPL-DM-BC: A Secure and Scalable FL Framework with Adaptive Privacy and Dynamic Masking, Blockchain, and XAI for the IoVs\
}

\author{     
    \IEEEauthorblockN{         
        Sathwik Narkedimilli\IEEEauthorrefmark{1},         
        Amballa Venkata Sriram\IEEEauthorrefmark{1},         
        Sujith Makam\IEEEauthorrefmark{1},         
        MSVPJ Sathvik\IEEEauthorrefmark{1},         
        Sai Prashanth Mallellu\IEEEauthorrefmark{2}
    }     
    \IEEEauthorblockA{         
        \IEEEauthorrefmark{1}Department of Computer Science, Indian Institute of Information Technology (IIIT) Dharwad\\         
        \{21bcs103, 21bcs008, 21bcs061\}@iiitdwd.ac.in, msvpjsathvik@gmail.com
    } 
    \IEEEauthorblockA{         
        \IEEEauthorrefmark{2}Department of Computer Science \& Engineering, Vardhaman College of Engineering\\         
        saiprashanth08@ieee.org
    } 
}

\maketitle

\begin{abstract}
The FAPL-DM-BC solution is a new FL-based privacy, security, and scalability solution for the Internet of Vehicles (IoV). It leverages Federated Adaptive Privacy-Aware Learning (FAPL) and Dynamic Masking (DM) to learn and adaptively change privacy policies in response to changing data sensitivity and state in real-time, for the optimal privacy-utility tradeoff. Secure Logging and Verification, Blockchain-based provenance and decentralized validation, and Cloud Microservices Secure Aggregation using FedAvg (Federated Averaging) and Secure Multi-Party Computation (SMPC). Two-model feedback, driven by Model-Agnostic Explainable AI (XAI), certifies local predictions and explanations to drive it to the next level of efficiency. Combining local feedback with world knowledge through a weighted mean computation, FAPL-DM-BC assures federated learning that is secure, scalable, and interpretable. Self-driving cars, traffic management, and forecasting, vehicular network cybersecurity in real-time, and smart cities are a few possible applications of this integrated, privacy-safe, and high-performance IoV platform.
\end{abstract}


\section{Introduction}

The explosion of the IoV is making possible faster, more effective communication, congestion management, and driverless vehicles. However, with IoV networks scaling exponentially, there are privacy, security, and scalability issues. Classic centralized learning systems don’t scale well with the distributed nature of IoV data and can result in data leakage, model poisoning, and real-time processing inefficiencies \cite{lv2020solving}. The current distributed FL solutions cannot meet the new requirements of dynamic IoV environments: in terms of privacy-usability, data provenance assurance, and scalable real-time execution \cite{ali2021efficient}.

To solve these issues, we present FAPL-DM-BC. This new Federated Learning system unifies Federated Adaptive Privacy-Aware Learning (FAPL), Dynamic Masking (DM), blockchain secure logging, and Explainable AI (XAI) in a cloud microservices system. The system dynamically adjusts privacy techniques like differential privacy or gradient masking, in real-time, depending on data sensitivity and environmental circumstances, for a perfect privacy-utility tradeoff \cite{da2021embedding}. Blockchain for IoVs also offers decentralized validation, provenance verification, and transparency via smart contracts to protect against data exploitation and model poisoning. In addition, the framework includes secure aggregation via FedAvg and Secure Multi-Party Computation (SMPC) for privacy-secure and fast global model updates. XAI, dual model feedback loop, and interpretable local predictions make the framework extremely robust and transparent \cite{bellucci2021towards}.

The FAPL-DM-BC model can be extended to many IoV domains. For cars — secure and decipherable learning — for the vehicle to make decisions in good faith, and data privacy. Traffic management and prediction The system enables vehicles and the infrastructure to work in real-time to flow more efficiently and eliminate congestion. It also improves vehicular network cybersecurity by recognizing and preventing network attacks using privacy protections. Smart City Integration, Secure and scalable data-based partnerships between IoV devices and urban infrastructure for effective resource management and smart transport networks: this is also a benefit of the framework.

Combining adaptive privacy, secure data management, and explainability, the FAPL-DM-BC platform is the all-in-one answer to the problems of IoV environments. It is modular and scalable allowing us to efficiently handle high-dimensional vehicle data and has great security and privacy protections. This architecture provides the intersection between distributed learning and real-world IoV usage so that autonomous vehicles, road systems, and smart cities can all function safely, intelligently, and collaboratively in real-time environments.

We present FAPL-DM-BC, a federated learning IoV system framework based on Federated Adaptive Privacy-Aware Learning (FAPL) and Dynamic Masking (DM) with privacy-utility tradeoffs. With blockchain to secure logging, decentralized validation, and irrevocable provenance, it prevents data theft and model poisoning. The architecture includes Secure Multi-Party Computation (SMPC), FedAvg for scale aggregation, and Model-Agnostic XAI with a dual model feedback loop for better precision and readability. Aimed at IoV use cases such as autonomous driving and traffic control, FAPL-DM-BC enables secure, scalable, and transparent IOV networks in real time.

\section{Preliminaries}

\subsection{Federated Adaptive Privacy-Aware Learning (FAPL)}

Federated Adaptive Privacy-Aware Learning (FAPL) \cite{zhao2020privacy} — a framework that will constantly update privacy policies in the context of federated learning to achieve the best possible data privacy/model utility trade-off. It includes contextual privacy tuning, and adjusting privacy-preserving techniques such as differential privacy and gradient masking in real-time based on data sensitivity, environment, and threat. FAPL allows IoV edge devices to dynamically improve their privacy with low model performance and is therefore the foundation for secure, scalable, and cost-effective federated learning in dynamic environments such as IoVs.

\subsection{Dynamic Masking (DM)}

Dynamic Masking (DM) \cite{anders2019spectral}: This privacy-preserving feature dynamically modifies the obfuscation of a model update (for example, gradients or logits) while transmitting. In contrast to static masking, DM considers the data’s sensitivity, environmental conditions, and threat to find the best masking strength. This flexibility gives good immunity against gradient-driven data leakage while still achieving good model convergence. DM can be useful in IoV systems where the data may be very sensitive and privacy as well accuracy are crucial.

\subsection{Secure Multi-Party Computation (SMPC)}

Secure Multi-Party Computation (SMPC) \cite{goldreich1998secure}: This is a cryptographic protocol that allows a group of people to cooccurringly solve a function without having their inputs revealed. SMPC in the FAPL-DM-BC model scheme is used in the global model aggregation for merging IoV edge device model updates without compromising the security of data. SMPC’s privacy-preserving aggregation keeps the federated learning process private and secure while preserving model precision. That’s why SMPC is a critical enabler of secure, scalable, and performant federated learning in IoV environments.

\subsection{Blockchain}

Blockchain \cite{shah2022blockchain} is a decentralized, fool-proof system where transactions or updates are recorded and verified without distortion or manipulation. For IoV federated learning, model updates are recorded on the blockchain, and verified by consensus algorithms such as PoS, and provenance information for data and model updates is tracked. Smart contracts do the validation and authorization for you automatically so that only approved updates are added to the global model. The decentralized aspect of blockchain makes the federated learning mechanism much more secure and trusted, thus, it is an integral part of the FAPL-DM-BC system.

\subsection{Explainable AI (XAI)}

By identifying the prediction process itself, Explainable AI (XAI) \cite{dwivedi2023explainable} gives machine learning models interpretation and transparency. In the FAPL-DM-BC architecture, XAI algorithms like SHAP, LIME, and Grad-CAM help IoV edge devices grasp local predictions and validate them. This openness is crucial for trust-building in applications such as driverless cars and traffic routing, where a decision has to be comprehensible and consistent. Then there is the double-model feedback loop, whereby predictions and explanations are constantly iteratively reworked to make models more accurate and accountable.

FAPL-DM-BC: is the first platform that combines Federated Adaptive Privacy-Aware Learning (FAPL) and Dynamic Masking (DM) with blockchain and Explainable AI (XAI) for the privacy, security, scalability, and interpretability problems of federated learning for IoV systems. It uses real-time adaptive privacy tuning instead of current solutions, which constantly updates privacy policies based on data risk and environmental risks, to ensure the best privacy-utility tradeoff. Blockchain and decentralized validation are combined for tamper-proof provenance tracking and decentralized validation and secure multi-party computation (SMPC) provide privacy-aware model aggregation. Also, the unique dual-model feedback loop based on model-independent XAI ensures local prediction accuracy and confidence, with input to global model updates through the weighted mean technique. This combination of the newest technologies and practices delivers a scalable, secure, and transparent federated learning framework for highly dynamic and sensitive IoV systems. 

\section{Literature Review}

The Internet of Vehicles (IoV) is a dynamic extension of the Internet of Things (IoT), enabling real-time data exchange among vehicles, devices, and environments. In \textit{Internet of Vehicles: Key Technologies, Network Model, Solutions and Challenges With Future Aspects} \cite{9300095}, the authors highlight cloud computing and big data analytics as essential for addressing scalability and heterogeneous network integration. Expanding on this, \textit{ A survey on internet of vehicles: applications, technologies, challenges, and opportunities} \cite{priyan2019survey} provides a broader discussion of IoT, IoV, and the Internet of Everything (IoE), identifying challenges like device interconnectivity, data security, and integration in critical applications such as healthcare.

Security challenges are a significant concern for IoV networks, as detailed in \textit{The Dark Side of The Internet of Vehicles: A Survey of the State of IoV and its Security Vulnerabilities} \cite{christensen2022darkinternetvehiclessurvey}. This paper emphasizes vulnerabilities to attacks such as DDoS and Sybil, advocating for blockchain technologies, encryption, and real-time monitoring to safeguard IoV systems. Together, these studies underline the importance of integrating scalable cloud-based frameworks with adaptive security measures to ensure the reliability and robustness of IoV networks.

The study `Explainable Federated Learning Scheme for Secure Healthcare Data` \cite{Zhao2024} introduces a framework that enhances model interpretability and transparency in federated learning for sensitive applications like healthcare. It emphasizes the importance of explainable AI in making complex models accessible and fostering trust among stakeholders. Its principles are highly adaptable to the Internet of Vehicles (IoV), where explainability is essential for ensuring safety and reliability in AI-driven systems.

Federated Learning (FL) has emerged as a promising approach to address privacy and security concerns in collaborative data training, but it faces challenges like protecting data privacy and intellectual property. One proposed solution is a framework that integrates privacy-preserving techniques with secure model protection \cite{Yang2023}, safeguarding both data privacy and the models’ intellectual property. Another solution, SHERPA \cite{10646830}, uses explainable AI (XAI) and federated learning to defend against data poisoning attacks, with the ability to detect malicious clients even when up to 80\% of participants are compromised, significantly improving robustness in federated environments. Additionally, the IFed framework enhances federated learning for the Power IoT with local differential privacy (LDP) \cite{lsngbhloigenowengoiwen}, offering privacy protection while maintaining model performance through dynamic privacy budget allocation.

Further addressing privacy and security concerns, blockchain-enabled federated learning (FL) systems are explored \cite{dfkjsbfkbnwekfnwekfnwken} as a means to integrate decentralized data management, ensuring secure and transparent communication in areas like healthcare and IoT. However, the integration of blockchain introduces high computational overhead, which could impact system efficiency. To enhance security in networked environments, a deep learning approach combining XAI and federated learning is proposed for intrusion detection systems \cite{sbfbwefbnweofnowenf}. This hybrid framework ensures model transparency while preserving privacy, although challenges exist in balancing explainability with accuracy and scaling federated learning across diverse network environments.

The FL-DECO-BC framework \cite{narkedimilli2024fldecobcprivacypreservingprovablysecure} presents another innovative approach for enhancing privacy, security, and provenance in Federated Learning, using decentralized oracles and blockchain to address issues related to trust and data integrity in Vehicular Ad-hoc Networks (VANETs). While the solution improves trust and data integrity, challenges remain in scalability and the overhead of integrating blockchain and oracles. Together, these approaches highlight the ongoing development of Federated Learning systems and their integration with other technologies to address emerging security, privacy, and scalability challenges.

\section{Proposed Framework}

The FAPL-DM-BC architecture is a novel FL architecture to solve privacy, security, and scalability issues in IoV. It uses adaptive privacy algorithms, blockchain-enabled tamper-proof logs, cloud microservices for secure collection, and Explainable AI (XAI) algorithms to provide robust, scalable, and readability. The framework offers Federated Adaptive Privacy-Aware Learning (FAPL) for flexible privacy, differential privacy, gradient masking, blockchain for secure provenance, and decentralized verification through smart contracts. An innovative dual-model feedback structure unites local prediction validation based on Model-Agnostic XAI models (i.e., SHAP, LIME, Grad-CAM) with global updates generated by FedAvg (Federated Averaging) \cite{qu2021federated} for interpretability, accuracy and confidence at the edge device level. The weighted mean fusion of local feedback and model update globality ensures consistency, flexibility, and ongoing improvement, in real-time data processing for low false positives. Its architecture makes federated learning in IoVs secure, explicable, and scalable, a single solution for distributed model training in high-vehicle-activity domains.

\subsection{The FAPL-DM-BC Workflow}

\begin{figure}
    \centering
    \fbox{\includegraphics[width=1\linewidth]{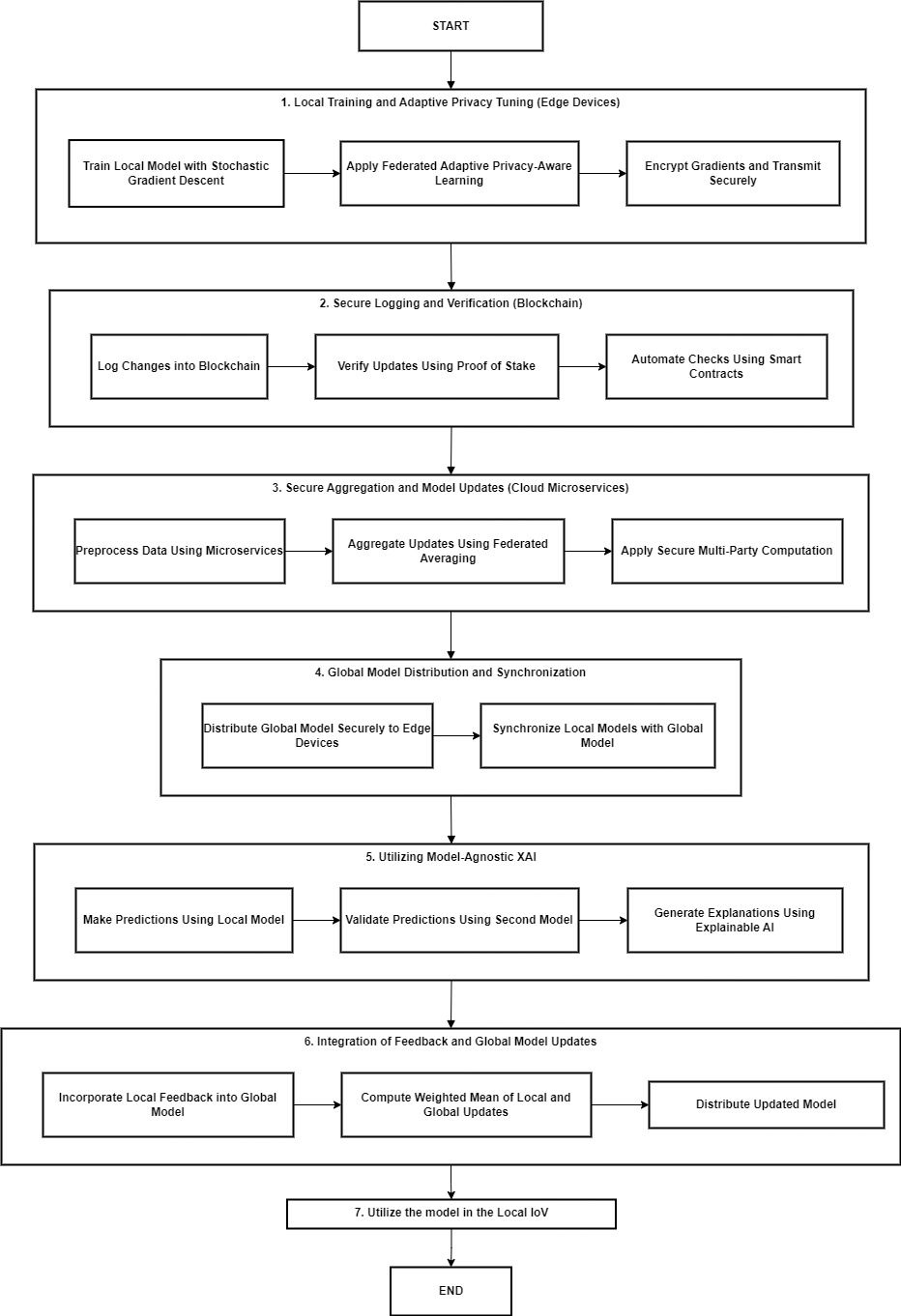}}
    \caption{The proposed FAPL-DM-BC Framework}
\end{figure}

\begin{enumerate}

\item \textbf{\textit{Local Training and Adaptive Privacy Tuning (Edge Devices)}}: 
    
Each IoV edge node (vehicle node) learns a local model from its dataset through stochastic gradient descent (SGD). It is trained with Federated Adaptive Privacy-Aware Learning (FAPL) which customizes privacy mechanisms based on data sensitivity, environment, and threat level in real-time. These tweaks include Adding differential privacy noise to gradients or parameters to mask confidential information and Gradient Masking to avoid gradient-based leakage. Contextual privacy tuning allows for proper tailoring of these privacy functions according to the specific state of each edge device using loss convergence rates, network security, model accuracy thresholds, etc. This makes sure that the privacy-utility trade-off is kept in check to protect data while still giving model performance.

After the local training, masked gradients or encrypted logits are encrypted and sent to the cloud microservices to be aggregated and processed. These changes are transmitted in encrypted communication, so they can’t be intercepted or tampered with while sending. FAPL’s adaptive privacy tuning allows the system to constantly tweak its privacy capabilities so that it’s resilient to new attacks. This step is important for keeping edge-generated data private and for a successful engagement in the federated learning process.

\item \textbf{\textit{Secure Logging and Verification (Blockchain)}}: 

Blockchain Interaction Microservice records model changes like the mask gradients or encrypt logits received from edge devices into a private blockchain. Each change is associated with a unique cryptographic hash, so data is always persistent and traceable. Nodes on the blockchain verify these updates via Proof of Stake (PoS) consensus to guarantee that they are true, fresh, and honest. This keeps malicious or outdated updates out of the aggregation pipeline. Blockchain smart contracts automate the verification and deny integrity checks \& privacy restrictions errors update. As the blockchain stores a record of each update in a tamper-proof manner, provenance tracking is secure in the federated learning process.

Validated updates are sent to the cloud microservices for secure aggregate. The blockchain is tamperproof which will provide transparency and confidence in the data storage, which prevents any attacks of model poisoning or unauthorized data modification. Adding blockchain into the federated learning process makes sure every update is tracked and verified before being released into the worldwide model. This is a critical phase to ensuring that the federated learning process is not compromised because it gives you a distributed source of data verification and provenance management.

\item \textbf{\textit{Secure Aggregation and Model Updates (Cloud Microservices)}}: 

The Secure Aggregation Microservice uses FedAvg (Federated Averaging) to aggregate the verified model updates from edge devices. FedAvg weights model parameters by local dataset size per device to provide fairness and equal contribution. The Data Preprocessing Microservice tidies up and formats the inbound updates before aggregation for heterogeneous data. Aggregation takes place via Secure Multi-Party Computation (SMPC) protocols, meaning the cloud can aggregate gradients or logits without exposing individuals, and edge devices won’t know who has contributed. This technique includes DM to dynamically control gradient obfuscation to ensure that the masking strength and model convergence are as close as possible.

Privacy Adjustment Microservice further filters the global model to add additional differential privacy noise to international parameters. It also guarantees that the final global model has strong privacy protection, protecting against a reverse engineer from one’s updates. When the aggregation is complete, the world model is registered on the blockchain to validate and track provenance. This is transparent and accountable from start to finish because every investment into the global model is logged and easily checkable. FedAvg and blockchain-powered provenance make for a powerful, effective, and private aggregation mechanism.

\item \textbf{ \textit{Global Model Distribution and Synchronization}}: 

Model Deployment Microservice, The model deploy microservice returns the gathered global model back to the associated vehicle nodes. The distributed global model is encrypted and sent to the edge devices over encrypted communications channels so that no enemy can intercept or alter the model while it is being distributed. Each vehicle node maps the local model to the global model, including collective knowledge of all participating nodes. This sync operation will keep the distributed fleet of cars in sync so that they can share in their joint training without compromising the confidentiality of their data.

The updated global model serves as the initial state for the next iteration of federated learning. This iterative process ensures continuous learning and adaptation, enabling the system to respond dynamically to changing data patterns, environmental conditions, and adversarial threats. By leveraging the FedAvg algorithm, blockchain-based validation, and cloud microservices, the FAPL-DM-BC framework ensures that the global model is securely aggregated, validated, and distributed, enabling robust and scalable federated learning in IoV environments. This step completes the federated learning cycle, ensuring that the system continuously evolves and improves while maintaining strong privacy and security guarantees.

\item \textbf{\textit{Utilizing Model-Agnostic XAI for Local Prediction, Validation, and Feedback Integration}}: 

All the IoV edge devices (vehicle nodes) use Model-Agnostic Explainable AI (XAI) to make the locally trained model’s predictions more understandable and trustworthy. Model 1 or the initial model, predicts based on the local model and produces explanations for its decisions using XAI tools like SHAP (Shapley Additive Explanations), LIME (Local Interpretable Model-Agnostic Explanations), or Grad-CAM. Such techniques are transparent in the sense that you know what characteristics shape the predictions so you can see what decisions will be made and then take action based on them. This capability is essential in an IoV environment, where knowing why things are done has implications for safety and productivity.

As part of making sure that these predictions and explanations are reliable and correct, a second model, Model 2, tests the results of Model 1. Model 2 determines whether or not the predictions are true and if the explanations match predicted decision paths. The process of validation is a local feedback loop where the models and corrections that Model 2 proposes can be applied to Model 1 to fix them locally. This process of continuous optimization improves the predictive accuracy as well as the explanations produced by the local device. With the dual-model model, not only does the edge device make valid predictions, but it also learns from them to evolve and increase its performance in real-time with the global update. \\

\item \textbf{ \textit{Integration of Feedback and Global Model Updates:}}: 

To maintain consistency and ensure the effectiveness of the federated learning framework, the feedback-derived updates from the local feedback loop (Models 1 and 2) are integrated with the global model updates through a weighted mean computation. This approach ensures that local refinements from the feedback loop are incorporated into the federated learning process while still leveraging the global knowledge aggregated across all participating edge devices. The integration process assigns a weight \( w_L \) to the local feedback updates, reflecting their significance in improving prediction accuracy and explanation reliability. Similarly, the global updates, derived from the FedAvg (Federated Averaging) algorithm, are assigned a weight \( w_G \), which is proportional to the size and diversity of the global dataset.

The final model updates are computed as:

\begin{center}
\[
\boxed{
\scalebox{0.9}{$
\text{Final Updates} = w_L \cdot x + w_G \cdot y
$}
}
\]
\end{center}

Where,

\begin{itemize}
    \item \( w_L \): Weight assigned to the local feedback updates, reflecting their significance in improving prediction accuracy and explanation reliability.
    \item \( w_G \): Weight assigned to the global model updates, derived from the FedAvg algorithm, proportional to the size and diversity of the global dataset.
    \item \( x \): Local feedback updates, incorporating corrections and refinements from the local feedback loop (Models 1 and 2).
    \item \( y \): Global model updates, representing the aggregated knowledge from all participating IoV devices.
\end{itemize}

The weights \( w_L \) and \( w_G \) can be calculated based on the \textit{contribution significance} and \textit{data volume}. For instance, \( w_L \) can be proportional to the accuracy improvement and explanation reliability from the local feedback loop, while \( w_G \) can be proportional to the size and diversity of the aggregated global dataset, ensuring balanced integration.

This weighted mean computation ensures that the final updates reflect the strengths of both the localized improvements from the feedback loop and the aggregated global knowledge from the federated learning process. By seamlessly combining local and global contributions, the framework ensures that the IoV system continuously evolves, with improved prediction and explanation quality at the local level while maintaining alignment with the global model and reducing the scope of false positives in the predictions. This integration enables robust, scalable, and interpretable federated learning for IoV environments.
    
\end{enumerate}

\subsection{Assumptions for the FAPL-DM-BC Framework}

The FAPL-DM-BC proposal assumes that the blockchain infrastructure for the IoV ecosystem is scalable enough to perform high-frequency, high-throughput operations including logging, validation, and consensus in real-time. The blockchain will have many model updates, cryptographic hashes, and provenance records sourced from a distributed network of edge devices with minimal latency or bottlenecks. It also depends on the efficiency of consensus algorithms like PoS to validate updates securely and fast. The blockchain’s scalability is the key to tamper-proof record recording without any issues and accommodating increasing edge devices in large-scale transportation networks.

It also assumes that the edge devices and the cloud system are compute-ready for cryptographic computations, data transmission, and federated model calculations. Edge nodes should perform privacy-preserving operations (differential privacy noise injection, gradient masking, encryption of logits or gradients before transmission). In the same way, cloud microservices must also be prepared for resource-demanding operations like Secure Multi-Party Computation (SMPC) for aggregate, differential privacy tuning for global models, and weighted merging of local feedback and global changes. Such computationally intensive functions are required for secure, privacy-safe, and fast-federated learning in the IoV.

The architecture also presupposes secure communication channels between edge nodes and cloud microservices to avoid the potential for data encapsulation or tampering. Also assumed is the access of XAI packages to edge devices that can perform local validation and feedback processing with dual-model architecture. Whether or not to weigh local feedback and global updates together is determined by quantifiable factors like many records, diversity, precision increase, and explanation stability. Such assumptions are critical for the scalability, readability, and security needed for the FAPL-DM-BC system to be able to scale with real-time vehicle situations and preserve the integrity of federated learning.

\subsection{BAN Logic Analysis for FAPL-DM-BC Framework}

BAN (Burrows-Abadi-Needham) Logic \cite{cohen2005completeness} is a formal framework to inspect and validate authentication protocol security. It is an effective way to reason about the views of individuals in a communication network so that trust, authenticity, and secrecy are not violated. BAN logic in the FAPL-DM-BC system is critical for verifying the correctness and safety of the communication between edge devices, blockchain nodes, and cloud microservices. BAN logic enables the framework to make sure that all of the major steps — model updates, provenance management, and feedback integration — are within the required security guarantees. The verification formally discovers weaknesses, makes protocols robust, and secures the framework’s security on IoV-sensitive data.

\subsection*{Assumptions}
\begin{itemize}
    \item \( P \): Represents an IoV edge device (vehicle node).
    \item \( C \): Represents the cloud microservices.
    \item \( B \): Represents the blockchain network (including blockchain nodes and smart contracts).
    \item \( K_{P-C} \): A secure shared key between the edge device \( P \) and cloud microservices \( C \).
    \item \( K_{P-B} \): A secure shared key between the edge device \( P \) and blockchain \( B \).
    \item \( K_{B-C} \): A secure shared key between blockchain \( B \) and cloud \( C \).
    \item \( M_P \): Masked gradients or encrypted logits transmitted from the edge device \( P \).
    \item \( M_{C} \): Aggregated global model updates processed by \( C \).
    \item \( H(M) \): A cryptographic hash function applied to \( M \).
    \item \( \#M \): Freshness of message \( M \).
    \item \( A \): An adversary attempting to intercept or tamper with the communication.
\end{itemize}

\subsection*{Goals}
\begin{itemize}
    \item \( P \mid \equiv C \mid \sim M_P \): \( P \) believes that \( C \) has received \( M_P \) as intended.
    \item \( C \mid \equiv B \mid \sim M_C \): \( C \) believes that \( B \) has securely logged and validated \( M_C \).
    \item \( B \mid \equiv P \mid \sim M_P \): \( B \) believes that \( P \) has sent \( M_P \) securely and without tampering.
    \item \( P \mid \equiv \#M_C \): \( P \) believes that the global model update \( M_C \) received is fresh and untampered.
\end{itemize}

\subsection*{Logical Steps}

\subsubsection* {\textbf{Step 1: Local Training and Privacy Tuning}}
\begin{itemize}
    \item \( P \mid \equiv \#M_P \): \( P \) believes that \( M_P \) (masked gradients or encrypted logits) is fresh.
    \item \( P \rightarrow C : M_P, K_{P-C} \): \( P \) securely transmits \( M_P \) to \( C \) using a shared key \( K_{P-C} \).
    \item \( C \mid \equiv P \mid \sim M_P \): \( C \) believes that \( M_P \) originates from \( P \).
    \item \( C \mid \equiv H(M_P) \): \( C \) verifies the integrity of \( M_P \) using a cryptographic hash \( H(M_P) \).
\end{itemize}

\subsubsection*{\textbf{Step 2: Secure Logging and Verification (Blockchain)}}
\begin{itemize}
    \item \( C \rightarrow B : M_P, K_{B-C} \): \( C \) sends \( M_P \) to \( B \) for logging and validation using the shared key \( K_{B-C} \).
    \item \( B \mid \equiv C \mid \sim M_P \): \( B \) believes that \( M_P \) was securely transmitted by \( C \).
    \item \( B \mid \equiv \#M_P \): \( B \) confirms the freshness of \( M_P \).
    \item \( B \rightarrow C : \text{Validation(M}_P\text{)}, K_{B-C} \): \( B \) sends the validation result back to \( C \).
    \item \( C \mid \equiv B \mid \sim \text{Validation(M}_P\text{)} \): \( C \) believes that \( B \) has securely validated \( M_P \).
\end{itemize}

\subsubsection*{\textbf{Step 3: Secure Aggregation and Model Updates}}
\begin{itemize}
    \item \( C \mid \equiv \#M_P \): \( C \) believes that all incoming \( M_P \) from edge devices are fresh.
    \item \( C \mid \equiv \text{FedAvg(M}_P\text{)} \): \( C \) securely aggregates the local updates \( M_P \) using the FedAvg algorithm.
    \item \( C \rightarrow B : M_C, K_{B-C} \): \( C \) sends the aggregated global model \( M_C \) to \( B \) for logging and validation.
    \item \( B \mid \equiv C \mid \sim M_C \): \( B \) believes that \( M_C \) was securely transmitted by \( C \).
    \item \( B \mid \equiv H(M_C) \): \( B \) verifies the integrity of \( M_C \) using a cryptographic hash.
\end{itemize}

\subsubsection*{\textbf{Step 4: Global Model Distribution and Synchronization}}
\begin{itemize}
    \item \( C \rightarrow P : M_C, K_{P-C} \): \( C \) securely transmits the global model \( M_C \) back to \( P \) using \( K_{P-C} \).
    \item \( P \mid \equiv \#M_C \): \( P \) verifies the freshness of \( M_C \) and believes it is the most recent global update.
    \item \( P \mid \equiv C \mid \sim M_C \): \( P \) believes that \( M_C \) originates from \( C \) and has not been tampered with.
\end{itemize}

\subsubsection*{\textbf{Step 5: Feedback and Weighted Integration}}
\begin{itemize}
    \item \( P \rightarrow C : \text{Feedback(M}_P\text{)}, K_{P-C} \): \( P \) sends local feedback updates to \( C \) for weighted integration.
    \item \( C \mid \equiv \text{Weighted(M}_P\text{, M}_C\text{)} \): \( C \) integrates local feedback \( M_P \) and global updates \( M_C \) using weighted computation \( w_L \) and \( w_G \).
    \item \( C \rightarrow B : \text{Updated(M}_C\text{)}, K_{B-C} \): \( C \) sends the final integrated model \( M_C \) to \( B \) for logging and validation.
    \item \( B \mid \equiv C \mid \sim \text{Updated(M}_C\text{)} \): \( B \) believes that the updated global model \( M_C \) is valid and securely transmitted.
\end{itemize}

BAN logic review of FAPL-DM-BC protocol assures that all interplay between IoV edge hardware, cloud microservices, and blockchain nodes is integrity, confidential, and authenticated. As long as the framework confirms all messages (e.g., local status updates, global models, comments) are new, unverified, and from trusted sources, secure communication and provenance tracking can be achieved. Cryptographic techniques such as public keys and hash functions protect data from being intercepted and tampered with, and blockchain logging guarantees impervious auditing and reporting. The symmetric integration of local feedback and global updates, in addition, allows for uniformity and trust across the federated learning workflow creating a secure, scalable, and robust system for IoV scenarios.

\section{Analysis of Proposed Framework}

FAPL-DM-BC solution can help overcome the challenge that IoV architecture brings by bringing together different top technologies to offer federated learning that is robust, scalable, and privacy-compliant. With Federated Adaptive Privacy-Aware Learning (FAPL) and Dynamic Masking (DM), privacy controls are adaptively tuned in real time according to environment and data value. This feature protects privacy at the same time as maintaining model performance, even when the driving environment is a fast-evolving one. Furthermore, blockchain for safe recording and decentralized verification provides provenance tracking that is tamper-proof and helps to prevent the problems of data corruption and model poisoning that are typical in IoV.

The most significant feature of the proposed architecture is that it facilitates continuous learning using Explainable AI (XAI). Using dual-model feedback loops where Model 1 makes predictions and explanations and Model 2 verifies them, the system continually updates its accuracy and interpretability. This double feedback mechanism allows IoV systems to adapt and move forward without the need for manual manipulation or training every time – thus lowering the overall cost of maintenance. XAI’s feature importance discovery not only makes the local models optimal but also helps identify shifts in the distribution of data, which keeps the system agile and reliable. This feature is particularly useful for applications such as road control and autonomous vehicles where decisions in real-time are needed.

The architecture also uses cloud microservices for scalability and modularity. As you will see, by dividing the secure aggregation, privacy adjustment, and global model synchronization work across the microservices, the system will be able to handle the computation load as efficiently as possible. Microservices allow dynamic resource allocation and the architecture scales following new IoV devices or larger amounts of data. The architecture is also highly available and fault tolerant ensuring seamless operation under dynamic vehicular conditions. In combination with Secure Multi-Party Computation (SMPC) and FedAvg, microservices ensure that the system can handle large-scale deployments without compromising privacy, security, or performance. Such a combination of powerful technologies makes FAPL-DM-BC flexible for new requirements of IoV systems such as smart cities, car cybersecurity, and smart traffic prediction.

\begin{table*}[htbp]
\caption{Comparison of FAPL-DM-BC Framework with Other Frameworks}
\label{tab:comparison}
\centering
\begin{adjustbox}{width=\textwidth}
\begin{tabular}{|p{5cm}|p{2.5cm}|p{2.5cm}|p{2.5cm}|p{2.5cm}|p{2.5cm}|p{2.5cm}|p{3cm}|}
\hline
\textbf{Attack} & \textbf{FL-PP-IPR \cite{Yang2023}} & \textbf{SHERPA \cite{10646830}} & \textbf{IFed \cite{lsngbhloigenowengoiwen}} & \textbf{BC/Smart Contracts + FL \cite{dfkjsbfkbnwekfnwekfnwken}} & \textbf{SNDL-XAI-FLID \cite{sbfbwefbnweofnowenf}} & \textbf{FL-DECO-BC \cite{narkedimilli2024fldecobcprivacypreservingprovablysecure}} & \textbf{FAPL-DM-BC (Proposed Framework)} \\
\hline
DoS Attacks (Sybil)           & Limited  & Limited   & Yes       & Yes       & Limited  & Yes       & Yes       \\
Spoofing Attacks              & Yes      & Limited   & Yes       & Limited   & Yes      & Yes       & Yes       \\
Tampering Attacks             & Yes      & Yes       & Yes       & Yes       & Yes      & Yes       & Yes       \\
Replay Attacks                & Limited  & -         & Yes       & Yes       & Limited  & Yes       & Yes       \\
Byzantine Fault Tolerance     & Yes      & Yes       & Yes       & Yes       & No       & Yes       & Yes       \\
Backdoor Attacks              & Yes      & Yes       & Yes       & Limited   & No       & Yes       & Yes       \\
Centralized Server Compromise & Limited  & No        & No        & Yes       & No       & Yes       & Yes       \\
Masquerading Attacks          & Yes      & Limited   & Yes       & Limited   & Yes      & Yes       & Yes       \\
Front Running Attack          & No       & -         & No        & -         & -        & -         & Yes       \\
Message Modification          & Yes      & Limited   & Yes       & Yes       & Yes      & Yes       & Yes       \\
Man-in-the-Middle Attack      & Yes      & -         & Yes       & Yes       & Limited  & Yes       & Yes       \\
Model Inversion               & Limited  & Limited   & Yes       & No        & Yes      & Yes       & Yes       \\
Eavesdropping                 & Yes      & Limited   & Yes       & Yes       & Limited  & Yes       & Yes       \\
Location Pinpointing          & Limited  & Limited   & Limited   & No        & Limited  & Limited   & Yes       \\
Anonymity                     & Yes      & -         & -         & Yes       & Yes      & Yes       & Yes       \\
Non-traceability and Impersonation & Yes  & Limited   & Yes       & Yes       & Limited  & Yes       & Yes       \\
Traffic Analysis              & Limited  & -         & Limited   & Limited   & Limited  & Limited   & Yes       \\
Data Poisoning                & Limited  & Limited   & Yes       & Yes       & Yes      & Yes       & Yes       \\
Side Channel Attack           & No       & No        & No        & Limited   & No       & No        & No        \\
Collusion Attack              & Limited  & Limited   & Yes       & Limited   & Limited  & Yes       & Yes       \\
\hline
\end{tabular}
\end{adjustbox}
\vspace{1em}
\par\textbf{Note:} In cases where specific details were not explicitly provided in the sources, the analysis has been interpreted subjectively. 
\end{table*}

\subsection{Anonymity}

The proposed FAPL-DM-BC protocol is anonymizing by using Federated Adaptive Privacy-Aware Learning (FAPL) and Dynamic Masking (DM) to mask private data in transmission. Differential privacy noise and gradient masking can also be used to mask the individual sources of data so that no individual edge device or data can be identified from the model. This anonymity is further bolstered by the decentralized, pseudonymous design of the blockchain: since cryptographic hashes rather than personal data are used to store updates, IoV participants remain anonymous during the learning phase.

\subsection{Non-traceability and Impersonation attacks}

Non-traceability and impersonation attacks are prevented by using secure logging and PoS consensus mechanisms based on blockchain technology. All the edge device’s data is hashed cryptographically and confirmed by blockchain nodes to be authentic and to deter malicious actors from pretending to be nodes. Automated validation with smart contracts, which has stringent integrity requirements to only accept valid updates. Dynamic Masking further eliminates traceability by dynamically obscuring data transmission patterns so that attackers cannot trace the data updates to individual edge devices.

\subsection{Message Modification Attack}

The system counters message modification attacks using encrypted communication channels for model update messages between edge nodes and cloud microservices. It is impossible to manipulate updates as they are transmitted because blockchain has a tamper-proof logging mechanism where the slightest modification renders the cryptographic hash of the previous update null. Secure Multi-Party Computation (SMPC) makes it possible to aggregate without divulging data and also gives you extra assurance against unauthorized data modifications in the process. 

\subsection{Replay attack and Man in the middle attack}

Attacks by replay and MTM are mitigated with blockchain validation and encrypted communication. Updates are authenticated by blockchain nodes through timestamps and cryptographic proofs so that old and duplicate updates are rejected. Encrypted channels encrypt data signals that can’t be intercepted or changed by the adversary. The dynamic checking of the dual-model feedback loop also ensures maliciously updated or replayed updates are spotted locally and won’t be in the global model and maintains the integrity of the federated learning.

\section{Conclusion}

In conclusion, FAPL-DM-BC tackles critical problems in IoV by offering an enterprise-grade, secure, and scalable Federated Learning (FL) architecture. Integrating Federated Adaptive Privacy-Aware Learning (FAPL) for real-time, real-time privacy tinkering for best privacy-utility tradeoffs and protecting personal information. Through blockchain secure logging and decentralized validation, provenance tracking is tamper-proof, and model updates are verified with integrity. The cloud microservices support effective and safe aggregated model updates via FedAvg and Secure Multi-Party Computation (SMPC). Second, the dual-model feedback process based on Model-Agnostic Explainable AI (XAI) makes local predictions and explanations much more interpretable and trustworthy in terms of ongoing optimization and model alignment with the global model. Because it combines local feedback with globally knowable knowledge through a weighted mean calculation, the model assures robust and readable federated learning. This is not only a breakthrough as FAPL-DM-BC overcomes privacy, data leakage, scalability, and lack of interpretability issues with legacy FL systems but opens up IoV applications to enable driverless cars, traffic management, vehicular cybersecurity, smart city, and more for a secure, privacy-preserving, high-performance IoV network future.

\section{Future Scope}

In the future of the FAPL-DM-BC system, in terms of making it capable of more dynamic IoV scenarios, while expanding its scalability, efficiency, and agility. : The next stage may be working to improve the blockchain infrastructure for even larger IoV networks at lower latency using lightweight consensus protocols like DAG or more sophisticated versions of PoS. What’s more, finding quantum-resistant cryptographic methods to ensure data transmission can help inoculate the system against future attacks. Combining more sophisticated XAI techniques (cause inference, counterfactual arguments, etc.) will increase the interpretation and reliability of predictions. Moreover, edge intelligence and federated reinforcement learning could be used to provide real-time decision-making and predictive functions for IoV applications such as adaptive traffic control or swarm-based automated vehicle control. Finally, cross-domain interoperability and integration with smart city systems can make the system a complete intelligent transportation platform which in turn will drive the innovation in IoV and federated learning solutions.

\bibliographystyle{IEEEtran}

\begin{thebibliography}{30}

\bibitem{lv2020solving} 
Z.~Lv, S.~Zhang, and W.~Xiu, 
``Solving the security problem of intelligent transportation system with deep learning,'' 
\emph{IEEE Transactions on Intelligent Transportation Systems}, vol.~22, pp.~4281--4290, 2020.

\bibitem{ali2021efficient} 
I.~Ali, Y.~Chen, N.~Ullah, R.~Kumar, and W.~He, 
``An efficient and provably secure ECC-based conditional privacy-preserving authentication for vehicle-to-vehicle communication in VANETs,'' 
\emph{IEEE Transactions on Vehicular Technology}, vol.~70, pp.~1278--1291, 2021.

\bibitem{da2021embedding} 
L.~Da Xu, Y.~Lu, and L.~Li, 
``Embedding blockchain technology into IoT for security: A survey,'' 
\emph{IEEE Internet of Things Journal}, vol.~8, pp.~10452--10473, 2021.

\bibitem{bellucci2021towards} 
M.~Bellucci, N.~Delestre, N.~Malandain, and C.~Zanni-Merk, 
``Towards a terminology for a fully contextualized XAI,'' 
\emph{Procedia Computer Science}, vol.~192, pp.~241--250, 2021.

\bibitem{zhao2020privacy} 
S.~Zhao, R.~Bharati, C.~Borcea, and Y.~Chen, 
``Privacy-aware federated learning for page recommendation,'' 
in \emph{2020 IEEE International Conference on Big Data (Big Data)}, 2020, pp.~1071--1080.

\bibitem{anders2019spectral} 
S.~Anders, D.~Noto, M.~Seilmayer, and S.~Eckert, 
``Spectral random masking: A novel dynamic masking technique for PIV in multiphase flows,'' 
\emph{Experiments in Fluids}, vol.~60, pp.~1--6, 2019.

\bibitem{goldreich1998secure} 
O.~Goldreich, 
``Secure multi-party computation,'' 
\emph{Manuscript. Preliminary Version}, vol.~78, pp.~1--108, 1998.

\bibitem{shah2022blockchain} 
K.~Shah, S.~Chadotra, S.~Tanwar, R.~Gupta, and N.~Kumar, 
``Blockchain for IoV in 6G environment: Review solutions and challenges,'' 
\emph{Cluster Computing}, vol.~25, pp.~1927--1955, 2022.

\bibitem{dwivedi2023explainable} 
R.~Dwivedi, D.~Dave, H.~Naik, S.~Singhal, R.~Omer, P.~Patel, B.~Qian, Z.~Wen, T.~Shah, G.~Morgan, et al., 
``Explainable AI (XAI): Core ideas, techniques, and solutions,'' 
\emph{ACM Computing Surveys}, vol.~55, pp.~1--33, 2023.

\bibitem{9300095} 
J.~Fadhil and Q.~Sarhan, 
``Internet of Vehicles (IoV): A Survey of Challenges and Solutions,'' 
in \emph{2020 21st International Arab Conference on Information Technology (ACIT)}, 2020, pp.~1--10.

\bibitem{priyan2019survey} 
M.~Priyan and G.~Devi, 
``A survey on internet of vehicles: applications, technologies, challenges and opportunities,'' 
\emph{International Journal of Advanced Intelligence Paradigms}, vol.~12, pp.~98--119, 2019.

\bibitem{christensen2022darkinternetvehiclessurvey} 
T.~Christensen, S.~Mandavilli, and C.~Wu, 
``The dark side of the Internet of vehicles: A survey of the state of IoV and its security vulnerabilities,'' 
arXiv:2211.05775, 2022. [Online]. Available: \url{https://arxiv.org/abs/2211.05775}.

\bibitem{Zhao2024} 
L.~Zhao, H.~Xie, L.~Zhong, et al., 
``Explainable federated learning scheme for secure healthcare data sharing,'' 
\emph{Health Information Science and Systems}, vol.~12, 2024. [Online]. Available: \url{https://doi.org/10.1007/s13755-024-00306-6}.

\bibitem{Yang2023} 
Q.~Yang, A.~Huang, L.~Fan, et al., 
``Federated learning with privacy-preserving and model IP-right-protection,'' 
\emph{Machine Intelligence Research}, vol.~20, pp.~19--37, 2023. [Online]. Available: \url{https://doi.org/10.1007/s11633-022-1343-2}.


\bibitem{10646830} 
C.~Sandeepa, B.~Siniarski, S.~Wang, and M.~Liyanage, 
``SHERPA: Explainable Robust Algorithms for Privacy-Preserved Federated Learning in Future Networks to Defend Against Data Poisoning Attacks,'' 
in \emph{2024 IEEE Symposium on Security and Privacy (SP)}, 2024, pp.~4772--4790.

\bibitem{lsngbhloigenowengoiwen} 
H.~Cao, S.~Liu, R.~Zhao, and X.~Xiong, 
``IFed: A novel federated learning framework for local differential privacy in Power Internet of Things,'' 
\emph{International Journal of Distributed Sensor Networks}, vol.~16, Article ID 1550147720919698, 2020. [Online]. Available: \url{https://doi.org/10.1177/1550147720919698}.

\bibitem{dfkjsbfkbnwekfnwekfnwken} 
Y.~Qu, M.~Uddin, C.~Gan, Y.~Xiang, L.~Gao, and J.~Yearwood, 
``Blockchain-enabled Federated Learning: A Survey,'' 
\emph{ACM Computing Surveys}, vol.~55, no.~11, 2022. [Online]. Available: \url{https://doi.org/10.1145/3524104}.

\bibitem{sbfbwefbnweofnowenf} 
K.~Fatema, M.~Anannya, S.~Dey, C.~Su, and R.~Mazumder, 
``Securing Networks: A Deep Learning Approach with Explainable AI (XAI) and Federated Learning for Intrusion Detection,'' 
in \emph{Data Security and Privacy Protection}, 2025, pp.~260--275.

\bibitem{narkedimilli2024fldecobcprivacypreservingprovablysecure} 
S.~Narkedimilli, R.~Kumar, N.~Kumar, R.~Reddy, and C.~P., 
``FL-DECO-BC: A Privacy-Preserving, Provably Secure, and Provenance-Preserving Federated Learning Framework with Decentralized Oracles on Blockchain for VANETs,'' 
arXiv:2407.21141, 2024. [Online]. Available: \url{https://arxiv.org/abs/2407.21141}.

\bibitem{qu2021federated} 
Z.~Qu, K.~Lin, Z.~Li, and J.~Zhou, 
``Federated learning’s blessing: FedAvg has linear speedup,'' 
in \emph{ICLR 2021 Workshop on Distributed and Private Machine Learning (DPML)}, 2021.

\bibitem{cohen2005completeness} 
M.~Cohen and M.~Dam, 
``A completeness result for BAN logic,'' 
in \emph{Proceedings of Methods for Modalities}, vol.~4, 2005.

\end{thebibliography}

\vspace{12pt}

\end{document}